%% file: PSCuniq_revised.tex
\documentclass[nofootinbib]{article}

\usepackage{graphicx}

\usepackage{rotating}

\usepackage{dcolumn}

\usepackage{amssymb}

\usepackage{amsmath}

\usepackage{bm}

\usepackage{listings}

\usepackage{fancyhdr}

\usepackage{wrapfig}

\usepackage{subfig}

\usepackage{jheppub}

\include{mydefs}

\newcommand{\C}{{\mathbb C}}

\title{Pure Lorentz spin connection theories and \\ uniqueness of General Relativity}

\author{Kirill Krasnov$^1$,} 
\emailAdd{kirill.krasnov@nottingham.ac.uk}
\author{Ermis Mitsou$^2$}
\emailAdd{ermitsou@physik.uzh.ch}

\affiliation{$^1$School of Mathematical Sciences, University of Nottingham, NG7 2RD, UK}
\affiliation{$^2$Center for Theoretical Astrophysics and Cosmology, Institute for Computational Science, University of Zurich, CH--8057 Z\"urich, Switzerland}

\abstract{General Relativity can be reformulated as a diffeomorphism invariant gauge theory of the Lorentz group, with Lagrangian of the type $f(F\we F)$, where $F$ is the curvature 2-form of the spin connection. A theory from this class with a generic $f$ is known to propagate eight degrees of freedom: a massless graviton, a massive graviton and a scalar. General Relativity in this formalism avoids extra degrees of freedom because the function $f$ is special and leads to the appearance of six extra primary constraints on the phase space variables. Our main new result is that there are other theories of the type $f(F\we F)$ that lead to six extra primary constraints. However, only in the case of GR the dynamics is such that these six primary constraints get supplemented by six secondary constraints, which gives the end result of two propagating degrees of freedom. This is how uniqueness of GR manifests itself in this ``pure spin connection" formalism. The other theories we discover are shown to give examples of irregular dynamical systems. At the linear level around (anti-)de Sitter space they have two degrees of freedom, as General Relativity, with the extra ones manifesting themselves only non-linearly.}

\begin{document}

\maketitle

\section{Introduction}

Lorentzian 4-dimensional vacuum General Relativity (GR) with a non-zero cosmological constant can be expressed in terms of the following pure spin connection action \cite{Mitsou:2019nlt} 
\beq \label{eq:SGR}
S_{\rm GR} := \frac{M^2}{\La} \int \[ \Tr \, \sqrt{\bm{z} \bm{X} \bm{z}} \]^2 \, , 
\eeq
where
\beq \label{eq:Omega}
X^{abcd} := \frac{1}{2}\, F^{ab} \we F^{cd} \, , \hspace{1cm}  \bm{z}_{abcd} := \frac{1}{2} \[ \( 1 + \be \) \( \et_{ac} \et_{bd} - \et_{ad} \et_{bc} \) + \( 1 - \be \) \vep_{abcd} \] \, ,  
\eeq
are to be interpreted as $6 \times 6$ matrices in antisymmetric pairs of Lorentz indices $[ab]$, and $F^{ab} := \ed A^{ab} + A^a_{\,\,\,c} \we A^{cb}$ are the curvature 2-forms of the spin connection 1-forms $A^{ab} \equiv - A^{ba}$. The object $\vep_{abcd}$ is the Levi-Civita symbol and the indices are displaced with the Minkowski metric $\et_{ab}$. As for the parameters, $M := \( 8 \pi G \)^{-1/2}$ is the reduced Planck mass, $\La$ is the cosmological constant and $\be^{-1}$ is the Immirzi parameter.\footnote{The trace of a matrix square root still has the cyclic property if these matrices are invertible \cite{Mitsou:2019nlt}, which is the case in \eqref{eq:SGR} for real $\be$, so one can also express the function entering the trace as $\sqrt{\bm{z}^2 \bm{X}}$.} In \eqref{eq:SGR} we make a slight abuse of notation by considering functions of 4-forms, but these are to be understood as functions of the corresponding scalar densities, which will always be denoted with a tilde 
\beq
\ph \equiv \ti{\ph} \, \ed^4 x \, , \hspace{1cm} \ti{\ph} := \frac{1}{4!}\, \vep^{\mu\nu\ro\si} \ph_{\mu\nu\ro\si} \, , \hspace{1cm}  \ed^4 x := \frac{1}{4!}\, \vep_{\mu\nu\ro\si} \, \ed x^{\mu} \we \ed x^{\nu} \we \ed x^{\ro} \we \ed x^{\si} \, , 
\eeq
so that $S_{\rm GR} \sim \int \ed^4 x \[ \Tr \, \sqrt{\bm{z} \ti{\bm{X}} \bm{z}} \]^2$. 

The action \eqref{eq:SGR} is a member of the class of theories first considered in \cite{TorresGomez:2009gs} and further studied in \cite{Krasnov:2011hi} in the context of the idea of gravity/Yang-Mills unification. This class is given by the actions of the form 
\beq \label{eq:Sf}
S = \int f \( \bm{X} \) \, ,
\eeq
where $f$ is a gauge-invariant function of the ${\mathfrak g}\otimes{\mathfrak g}$ valued matrix $\bm{X}$, where ${\mathfrak g}$ is the Lie algebra of some gauge group $G$, and $\bm{X}$ is the product of the two copies of the Lie algebra valued curvature 2-form $X\sim F\wedge F$. The Lagrangian is a 4-form when the function $f$ is homogeneous of degree one in $\bm{X}$
\beq \label{eq:Ldeg1}
f \( \al \bm{X} \) \equiv \al f \( \bm{X} \) \, .
\eeq
This is necessary for the action \eqref{eq:Sf} to be invariant under the full diffeomorphism group. It is natural to refer to theories of this type as {\it diffeomorphism invariant gauge theories}. 

In \cite{Krasnov:2011pp} it was shown that GR can be formulated as such a diffeomorphism invariant gauge theory with $G = {\rm SO}(3)$ for Euclidean signature and $G = {\rm SO}(3,\C)$ for Lorentzian signature, with the latter case requiring extra reality constraints. Remarkably, for these 3-dimensional groups all the members of the class \eqref{eq:Sf} share the same degree of freedom (DoF) count as GR, as shown perturbatively in \cite{Krasnov:2011up} and fully non-linearly in their preceding canonical/BF formulations \cite{Bengtsson:1990qg, Bengtsson:1991bq, Krasnov:2006du, Krasnov:2007cq, Bengtsson:2007zzd}. In the Lorentzian case, where reality constraints are needed to obtain two real DoF (instead of complex), it was recently shown that compatible reality constraints only seem to exist for the case of GR \cite{Krasnov:2020zfi}, thus excluding the existence of viable physical modified gravity theories of that kind. 

Theories of the type \eqref{eq:Sf}, but for the real Lorentz group $G = {\rm SO}(1,3)$, were first considered in the context of gravity/Yang-Mills unification in \cite{Smolin:2007rx} in their BF formulation. In \cite{Alexandrov:2008fs} they were shown to generically propagate eight DoF. In \cite{Speziale:2010cf} these DoF were interpreted as those of a massless graviton, a massive graviton and a scalar, i.e. the typical spectrum of a bi-gravity theory. Works \cite{Beke:2011mu,Beke:2011dp} considered special members of this class of theories (but in the Euclidean signature corresponding to $G = {\rm SO}(4)$) that describe only the massless graviton and a scalar. 

The work \cite{Mitsou:2019nlt} (see Section 8) established that in the case of $G = {\rm SO}(1,3)$ the Lorentzian signature forces one of the two gravitons to be a ghost, i.e. a particle of negative kinetic energy, meaning that the generic theory of the class \eqref{eq:Sf} is not physical. One must therefore restrict their attention to the cases where the second graviton is absent, such as the scalar-tensor theories discussed in \cite{Beke:2011dp}. Although \cite{Beke:2011dp} works with Euclidean signature, the analysis can be translated straightforwardly to the Lorentzian case, which makes it clear that there is enough parameter freedom to obtain ghost-free scalar-tensor theories. 

General Relativity in this formalism \eqref{eq:SGR} corresponds to a theory for which the ``defining" function $f$ is very special. Thus, a more detailed analysis shows that for this choice of $f$ there appear extra primary constraints on the phase space variables of the theory. There are six constraints of this sort, known in the literature as the ``simplicity" or ``metricity" constraints. It is then a very natural question to ask if there are any other members from the same theory class that also have extra primary constraints. Indeed, there could be other interesting (and viable) theories in the class \eqref{eq:Sf} with $G = {\rm SO}(1,3)$ apart from GR, and the first step in uncovering such theories would be to establish the appearance of the primary constraints. The main new observation in this paper is that, indeed, other such theories do exist: a one-parameter deformation of GR \eqref{eq:defGRS} and a disconnected one-parameter family \eqref{eq:detS}. The presence of these constraints means that these theories too propagate less degrees of freedom than the generic member of the class \eqref{eq:Sf}. Moreover, we find that their linearisation around the maximally symmetric background, that is (anti-)de Sitter space ((A)dS), has the same DoF count as GR. One could therefore hope that these theories have the same spectrum as GR also non-linearly, thus invalidating the GR uniqueness results and giving new examples of diffeomorphism invariant theories describing interacting gravitons. 

This does not happen, and the underlying reasons are very instructive. It turns out that only in the case of GR the dynamics is such that the evolution of the primary constraints leads to six secondary constraints. These together then form a second-class pair which provides the required reduction in the number of DoF. For the theories that have primary constraints, but are distinct from GR, we find that no secondary constraints arise for {\it generic} field configurations, meaning they have more DoF than GR. However, since these theories have two DoF at the linear level around (A)dS, the obstruction to six secondary constraints arises only through interactions around that background. This is an example of {\it irregular dynamical systems} and, in particular, the phenomenon of ``(non-linear) field activation'' \cite{Yo:2001sy, Deser:2012ci}, i.e. linearly absent DoF that are ``activated'' by interactions. It is also worth noting that the theories we present here constitute qualitatively new examples of irregularity. Indeed, typically field activation is due to gauge symmetries at the linear level that are broken when including interactions, i.e. less first-class constraints non-linearly (and also possibly second-class).\footnote{For a recent example see the theories studied in \cite{Alexandrov:2021qry}, where a linearised conformal symmetry is broken by non-linearities, thus leading to the activation of one DoF. Poincar\'e gauge theory also exhibits this phenomenon quite generically \cite{Yo:2001sy}. Another popular example are partially massless graviton theories, where the linear theory on (A)dS propagates only the $\{ \pm 1, \pm 2 \}$ helicities of a massive graviton thanks to an Abelian U(1) gauge symmetry \cite{Deser:1983mm,Deser:2001us}, but this symmetry cannot survive non-linearly \cite{deRham:2013wv, Garcia-Saenz:2014cwa} unless the theory includes ghosts \cite{Joung:2014aba, Boulanger:2019zic}. In that case the prime example is conformal gravity, where the spectrum around (A)dS is a massless graviton and a partially-massless ghost graviton \cite{Maldacena:2011mk, Deser:2012qg, Kluson:2013hza} and the aforementioned U(1) symmetry is the combination of a conformal transformation and a diffeomorphism.} In contrast, here we have primary constraints that are the half of a second-class pair in the linear theory, but fail to produce the other (secondary) half non-linearly, becoming self-conjugate instead. Interestingly, the particular scenario that is the subject of this paper is only possible in a specific dimension. Our results thus show that the point describing GR has quite exotic neighbours in the theory space. This offers a novel perspective on the uniqueness of the theory.

This situation is ultimately due to the fact that the obstruction to secondary constraints depends on field values that can vanish in some limits (e.g. linearisation) or at special, yet accessible points of the phase space. As a result, the Dirac algorithm becomes ambiguous, as it can bifurcate into different constraint chains depending on the points of phase space, a phenomenon known as ``constraint bifurcation'' \cite{Yo:2001sy, Deser:2012ci}. In our case, we will see that the obstructing field combination is non-zero and non-degenerate generically, meaning that there are no secondary constraints in general. At the end of the Dirac algorithm, one is therefore left with five DoF, i.e. three more than in GR. Since the special theories we identify here are the most general set with six extra primary constraints, our results allow us to characterise GR as the only theory from the class \eqref{eq:Sf} with strictly two DoF.

The mismatch between the DoF of a theory and the ones of its linearisation is generally considered a disqualifying pathology, at least for the background solution around which one linearizes. Indeed, it implies that the concerned modes are infinitely strongly coupled around that background, since the absence of linear terms in the action makes finite interactions infinitely more intense. In particular, this invalidates any perturbative/effective field theory approach. Another potential issue in our case is that part of the extra DoF with respect to GR are associated to the ghost graviton in the generic theory. In the absence of a linear kinetic term around (A)dS, the impact of such an excitation on stability is ambiguous, but it will be present linearly around more general backgrounds, so it seems hard to avoid a ghost issue in general. Finally, the fact that the DoF count depends on regions of phase space might imply that the Cauchy problem is generically not well-defined. Still, the ultimate utility of the new theories described in this paper can only be decided by further dedicated study.      

Our results are of relevance for the spin foam approach to quantum gravity \cite{Oriti:2001qu,Perez:2003vx}. This approach proceeds starting with the state sum models of ${\rm SO}(1,3)$ topological BF theory, and then modifying these by adding constraints on configurations to be summed over that are designed to mimic the primary ``simplicity" constraints on the $B$ field. The two most popular models that arise this way are \cite{Engle:2007wy,Freidel:2007py}. The secondary constraints that are also present in the continuous version of the theory are often ignored in the discrete spin foam model setting. It has been long argued by Alexandrov, see in particular \cite{Alexandrov:2008da}, that the secondary constraints should also be dealt with explicitly in the path integral. The new examples of gravitational theories that we discover in this paper, which have the primary constraints but not the secondary ones, add an interesting new dimension to this discussion. Indeed, our results imply that GR is the only theory (in a certain large class) that possesses {\bf both} primary and secondary constraints, but there are other theories if the requirement of having the secondary constraints is dropped. This makes it clear that one needs both primary and secondary constraints to pinpoint GR, which adds additional support to the arguments of \cite{Alexandrov:2008da}. 

The paper is organized as follows. In section \ref{sec:defs} we briefly introduce some notation and conventions, especially for the case of most interest that is the group SO(1,3). In section \ref{sec:genprop} we lay out the tools required for our purposes. Most of the discussion here is valid for any gauge group $G$. In section \ref{sec:vaclin} we describe the linearised theory around the (A)dS solution in detail. In section \ref{sec:GRuniq} we identify the aforementioned special theories and perform a fully non-linear canonical analysis that demonstrates their irregular dynamics. Finally, in section \ref{sec:conclusion} we provide some concluding remarks.

\section{Notation \& conventions} \label{sec:defs}

Consider a semi-simple Lie group $G$, a set of connection 1-forms $A^{\fa}$ and the associated curvature 2-forms
\beq
F^{\fa} := \ed A^{\fa} + \frac{1}{2}\, f^{\fa}_{\,\,\,\fb\fc} A^{\fb} \we A^{\fc} \, ,
\eeq 
where the $f^{\fa}_{\,\,\,\fb\fc}$ are the structure coefficients of the associated Lie algebra $\frak{g}$. The algebra indices $\fa,\fb,\fc,\dots$ are then displaced using the killing form $\ka_{\fa\fb} \propto f^{\fc}_{\,\,\,\fd\fa} f^{\fd}_{\,\,\,\fc\fb}$ and, since the group is semi-simple, one can choose the generator basis such that $\ka_{\fa\fb}$ is diagonal with $\pm 1$ entries. 

Many of the computations in this paper will hold for arbitrary gauge group $G$, but the case of direct interest will be the Lorentz group SO(1,3). In that case the algebra indices can be expressed as antisymmetric pairs of Lorentz indices $[ab]$, the latter being always contracted with a combinatoric $1/2$ factor to avoid counting each independent component twice, as is assumed in \eqref{eq:SGR} for instance. The Killing form and structure coefficients are
\bea
\ka_{[ab][cd]} & \equiv & \et_{ac} \et_{bd} - \et_{ad} \et_{bc} \, , \\
f_{[ab][cd][ef]} & \equiv & \et_{ac} \ka_{[bd][ef]} - \et_{ad} \ka_{[bc][ef]} - \et_{bc} \ka_{[ad][ef]} + \et_{bd} \ka_{[ac][ef]} \, ,
\eea
and we choose the sign conventions 
\beq
\et \equiv {\rm diag}(-1,1,1,1) \, , \hspace{1cm} \vep_{0123} \equiv 1 \, ,
\eeq
so, in particular,
\beq
\ka \equiv {\rm diag}(-1,-1,-1,1,1,1) \, , 
\eeq
We can then define the invariant identity $\bm{1}$ and dual $\star$ operators on the algebra  
\beq
\bm{1}^{[ab]}_{\,\,\,\,\,\,\,\,[cd]} := \ka^{[ab]}_{\,\,\,\,\,\,\,\,[cd]} \, , \hspace{1cm} \star^{[ab]}_{\,\,\,\,\,\,\,\,[cd]} := \vep^{ab}_{\,\,\,\,\,\,cd} \, ,
\eeq
where the latter acts as an imaginary unit since 
\beq \label{eq:starprop}
\star^2 \equiv -\bm{1} \, .
\eeq
Then the following invariant combinations
\beq \label{eq:zalbe}
\bm{z} := \al \bm{1} + \be \star \, , \hspace{1cm} \al,\be \in \Rs \, ,
\eeq
such as the one appearing in \eqref{eq:Omega}, satisfy the algebraic relations of the complex numbers. In particular, they are invertible and we can decompose them into a ``modulus'' and ``phase'' $\bm{z} \equiv r e^{\te \star}$. In what follows the bold letters $\bm{z}$ and $\bm{w}$ will always denote such invariant operators with real coefficients. Finally, the interested reader can look at section 2 of \cite{Mitsou:2019nlt} for how the operators $\bm{1}$ and $\star$ can be used to decompose any matrix $\bm{M}$ into its irreducible parts under the Lorentz group.

\section{Diffeomorphism invariant gauge theories} \label{sec:genprop}

This section reproduces and extends results contained in \cite{TorresGomez:2009gs}, \cite{Krasnov:2011hi}.

\subsection{Lagrangian formulation}

Consider the type of action \eqref{eq:Sf}, but now for an arbitrary gauge group
\beq \label{eq:gen}
X^{\fa\fb} := \frac{1}{2}\, F^\fa \we F^\fb \, .
\eeq
The equations of motion read
\beq \label{eq:EOM}
\ced \psi_{\fa\fb} \we F^{\fb} = 0 \, , 
\eeq
where we have defined the 0-forms
\beq \label{eq:psidef}
\psi_{\fa\fb} := \frac{\pa f}{\pa X^{\fa\fb}} \, , 
\eeq
$\ced$ is the exterior covariant derivative and we have used the Bianchi identity $\ced F^{\fa\fb} \equiv 0$. As we shall see, this $\psi_{\fa\fb}$ matrix is the generalization to arbitrary group of the $\psi_{ij}$ matrix considered in \cite{Krasnov:2020zfi} for the case of $G = {\rm SO}(3,\Cs)$. To maintain a consistent notation, here too we will denote the inverse matrix components by $X_{\fa\fb}$ and $\psi^{\fa\fb}$, respectively, i.e. 
\beq
X_{\fa\fc} X^{\fc\fb} \equiv \de^{\fa}_{\fb} \, , \hspace{1cm} \psi^{\fa\fc} \psi_{\fc\fb} \equiv \de^{\fa}_{\fb} \, ,
\eeq
therefore paying attention to index positions. Note that $X^{\fa\fb}$ (and therefore $\psi_{\fa\fb}$) is invertible only if the group dimension $D$ is at most six, because it is made of six $D$-vectors (see \cite{Mitsou:2019nlt} or \eqref{eq:XofKB}). Since $D = 6$ is the case of most interest here, we will assume invertibility to simplify some derivations and expressions, meaning that not all of them presented here generalize as such to $D > 6$. Following again \cite{Krasnov:2020zfi}, when using a compact matrix notation, $\bm{X}$ and $\bm{\psi}$ will denote the upper index matrices $X^{\fa\fb}$ and $\psi^{\fa\fb}$, respectively, so that $\bm{X}^{-1}$ and $\bm{\psi}^{-1}$ denote the lower index ones $X_{\fa\fb}$ and $\psi_{\fa\fb}$, respectively. Contraction between indices of the same type are then understood to be performed using the Killing form $\ka_{\fa\fb}$ and $\ka^{\fa\fb}$. 

We can now note that the addition of a topological term to the Lagrangian 
\beq \label{eq:fshift}
f \to f + z_{\fa\fb} X^{\fa\fb} \, ,
\eeq
where $z_{\fa\fb}$ is some invariant matrix under the group action, does not change the classical physics and corresponds to shifting $\psi_{\fa\fb}$
\beq \label{eq:Xishift}
\psi_{\fa\fb} \to \psi_{\fa\fb} + z_{\fa\fb} \, .
\eeq
Next, by differentiating \eqref{eq:Ldeg1} with respect to $X^{\fa\fb}$, we find that the $\psi_{\fa\fb}$ are homogeneous functions of $\bm{X}$ of degree zero
\beq \label{eq:Labdeg0}
\psi_{\fa\fb} \( \al \bm{X} \) \equiv \psi_{\fa\fb} \( \bm{X} \) \, ,
\eeq
and therefore a set of scalars under diffeomorphisms. On the other hand, taking the derivative with respect to $\al$ in \eqref{eq:Ldeg1} and then setting $\al = 1$ we get
\beq \label{eq:zeroLeg}
\psi_{\fa\fb} X^{\fa\fb} \equiv f \, ,
\eeq
which says that the Legendre transform of $f$ with respect to $X^{\fa\fb}$ is zero. Further differentiating \eqref{eq:zeroLeg} with respect to $X^{\fa\fb}$ we find
\beq \label{eq:LabcdXcd0}
H_{\fa\fb,\fc\fd} X^{\fc\fd} \equiv 0 \, , \hspace{1cm} H_{\fa\fb,\fc\fd} := \frac{\pa^2 f}{\pa X^{\fa\fb} \pa X^{\fc\fd}} \, ,
\eeq
so the Hessian matrix $H_{\fa\fb,\fc\fd}$ is not invertible, since it has an eigenvector $X^{\fa\fb}$ with zero eigenvalue. Thus, the relation between Legendre-conjugate variables $\bm{\psi} = \bm{\psi}(\bm{X})$ cannot be uniquely inverted. The ambiguity lies in an arbitrary multiplicative factor, because if $\bm{X} = \bm{X}(\bm{\psi})$ is some solution, then so is $\al \bm{X}(\bm{\psi})$, thanks to \eqref{eq:Labdeg0}. The inversion is therefore unique only up to an arbitrary 4-form $\ph$
\beq \label{eq:XofLgen}
\bm{X} = \ph \hat{\bm{X}}(\bm{\psi}) \, ,
\eeq
and the normalization of $\hat{\bm{X}}$ is free to choose, as any modification can be reabsorbed in $\ph$. The presence of $\ph$ can also be understood by the fact that one cannot produce 4-forms $X^{\fa\fb}$ out of the 0-forms $\psi_{\fa\fb}$ alone. Now if we insert \eqref{eq:XofLgen} inside \eqref{eq:zeroLeg} and use \eqref{eq:Ldeg1} and \eqref{eq:Labdeg0}, we get
\beq \label{eq:ticHdef}
\psi_{\fa\fb} X^{\fa\fb}(\bm{\psi}) - f(\bm{X}(\bm{\psi})) \equiv \ph \cH(\bm{\psi}) \, ,
\eeq
for the following function $\cH$ of $\psi_{\fa\fb}$
\beq \label{eq:HtofL}
\cH(\bm{\psi}) := \psi_{\fa\fb} \hat{X}^{\fa\fb}(\bm{\psi}) - f(\hat{\bm{X}}(\bm{\psi})) \, ,
\eeq
and thus \eqref{eq:zeroLeg} becomes a scalar identity for $\psi_{\fa\fb}$
\beq \label{eq:HofX0}
\cH(\bm{\psi}) \equiv 0 \, .
\eeq
This can be understood by the fact that the number of independent components on both sides of \eqref{eq:XofLgen} must match, i.e. if we have a free function $\ph$ on the right-hand side, then $\bm{\psi}$ must obey a constraint. Taking the general variation of \eqref{eq:HtofL} we also verify the standard properties of the Legendre transform that are $\cH \equiv \cH(\bm{\psi})$ and
\beq \label{eq:hatXofH}
\hat{X}^{\fa\fb} = \frac{\pa \cH}{\pa \psi_{\fa\fb}} \, .
\eeq
One can then note that, given \eqref{eq:HofX0}, equations \eqref{eq:EOM}, \eqref{eq:XofLgen}, and \eqref{eq:hatXofH} are nothing but the equations of motion of the action
\beq \label{eq:SH}
S = \int \[ \frac{1}{2}\, \psi_{\fa\fb} F^{\fa} \we F^{\fb} - \ph \cH (\bm{\psi}) \] \, ,
\eeq
where $\ph$ and $\bm{\psi}$ are independent auxiliary fields. This is the Legendre transform of \eqref{eq:Sf} with respect to $\bm{X}$. We see that $\ph$ appears as a Lagrange multiplier imposing $\cH = 0$, which now appears as an equation of motion, rather than an identity. The form \eqref{eq:SH} is the generalization of equation (2.11) of \cite{Krasnov:2020zfi} to arbitrary group $G$. Here we have reached this form by starting from the pure-connection action, whereas in \cite{Krasnov:2020zfi} it was derived by integrating out the $B$ field in the $BF$ formulation.

As a concrete example, consider Lorentzian GR \eqref{eq:SGR}. We have
\beq \label{eq:LGR}
f_{\rm GR}(\bm{X}) = M^2 \La^{-1} \[ \Tr \, \sqrt{\bm{X}_{\bm{z}}}  \]^2 \, , \hspace{1cm} \bm{X}_{\bm{z}} := \bm{z} \bm{X} \bm{z} \, , \hspace{1cm} \bm{z} := \frac{1}{2} \[ \( 1 + \be \) \bm{1} + \( 1 - \be \) \star \] \, ,
\eeq
so that \eqref{eq:psidef} gives
\beq \label{eq:XofLex}
\bm{\psi}^{-1}_{\rm GR}(\bm{X}) = M^2 \La^{-1} \[ \Tr \, \sqrt{\bm{X}_{\bm{z}}} \] \bm{z} \( \bm{X}_{\bm{z}} \)^{-1/2} \bm{z}   \, .
\eeq
One possible inversion \eqref{eq:XofLgen} is
\beq \label{eq:XofLGR}
\hat{\bm{X}}_{\rm GR}(\bm{\psi}) = \La \, \frac{\bm{\psi} \bm{z}^2 \bm{\psi}}{\Tr \( \bm{z} \bm{\psi} \bm{z} \)} \, ,
\eeq
so that, inserting this inside \eqref{eq:HtofL}, we obtain 
\beq \label{eq:HGR}
\cH_{\rm GR}(\bm{\psi}) := \La - M^2\, \Tr \( \bm{z} \bm{\psi} \bm{z} \) \, ,
\eeq
which is indeed zero when one uses \eqref{eq:XofLex}. We also verify that \eqref{eq:hatXofH} holds when $\cH_{\rm GR} = 0$ is taken into account.

\subsection{Canonical formulation} \label{sec:caness}

Let us now foliate the manifold $x^{\mu} = \{ t, x^{\al} \}$, i.e. using Greek letters from the beginning of the alphabet $\al,\be,\ga,\dots$ to denote spatial diffeomorphism indices and denoting by $\Si_t$ the $t = {\rm constant}$ hypersurfaces. We define the ``kinetic'' and ``magnetic'' fields
\beq
K_{\al}^\fa := F^\fa_{t\al} \, , \hspace{1cm} B^{\fa\al} := \frac{1}{2}\, \vep^{\al\be\ga} F_{\be\ga}^\fa \, ,
\eeq
so that 
\beq \label{eq:XofKB}
X^{\fa\fb} \equiv K_{\al}^{(\fa} B^{\fb)\al} \, \ed^4 x \, , \hspace{1cm} \ed^4 x := \frac{1}{4!} \, \vep_{\mu\nu\ro\si}\, \ed x^{\mu} \we \ed x^{\nu} \we \ed x^{\ro} \we \ed x^{\si} \, .
\eeq
To derive the canonical action, we can either Legendre-transform the original action \eqref{eq:Sf}, or we can manipulate its alternative formulation \eqref{eq:SH} which is already linear in time-derivatives. In the latter case the independent $\psi_{\fa\fb}$ field contains the conjugate momentum information, but for a gauge group with $D > 3$ it also contains extra independent components. These must therefore be properly disentangled and integrated out, which is more complicated than the approach we will use here. So we start by computing the conjugate momenta of \eqref{eq:Sf}, or ``electric'' fields, 
\beq \label{eq:Edef}
E^{\al}_\fa := \frac{\pa f}{\pa \dot{A}^\fa_{\al}} = \frac{\pa f}{\pa K^\fa_{\al}} = \psi_{\fa\fb} B^{\fb\al} \, ,
\eeq
where $\psi_{\fa\fb}$ is given by \eqref{eq:psidef}. As we will see, at least for the groups of interest here, $\bm{\psi}$ will always be expressible in terms of the electric and magnetic fields 
\beq \label{eq:XimustEB}
\bm{\psi} \equiv \bm{\psi}(E,B) \, ,
\eeq
and therefore in terms of the canonical variables. Next, if we define
\beq
\te^\fa := - A_t^\fa \, ,
\eeq 
then the Legendre transform of the action with respect to $\dot{A}_{\al}^a$ reads
\beq \label{eq:Ltrans}
E_\fa^{\al} \dot{A}^\fa_{\al} - \ti{f} \equiv E_\fa^{\al} K^\fa_{\al} - \ti{f} - E_\fa^{\al} \na_{\al} \te^\fa = \psi_{\fa\fb} K^\fa_{\al} B^{\fb\al} - \ti{f} - E_\fa^{\al} \na_{\al} \te^\fa \equiv \psi_{\fa\fb} \ti{X}^{\fa\fb} - \ti{f} - E_\fa^{\al} \na_{\al} \te^\fa \, ,
\eeq
where $\na_{\al}$ is the covariant derivative with respect to gauge transformations on $\Si_t$
\beq
\na_{\al} \te^\fa := \pa_{\al} \te^\fa + f^\fa_{\,\,\,\fb\fc} A_{\al}^\fb \te^\fc \, .
\eeq
Note that we have not imposed \eqref{eq:zeroLeg}, because that equation holds only when the momenta are on-shell \eqref{eq:Edef}, which is no longer the case in the canonical formalism, i.e. they are independent variables. Instead, one must now use \eqref{eq:ticHdef} to finish the computation \eqref{eq:Ltrans}
\beq
E_\fa^{\al} \dot{A}^\fa_{\al} - \ti{f} = \ti{\ph} \cH(\bm{\psi}) - E_\fa^{\al} \na_{\al} \te^\fa \, .
\eeq  
We now see that $\ph$ plays the role of a Lagrange multiplier imposing the Hamiltonian phase space constraint $\cH(\bm{\psi}) = 0$, thus recovering the condition \eqref{eq:zeroLeg} on-shell. As for the $\te^\fa$, they impose as usual the Gauss constraint associated with gauge invariance
\beq
\cG_\fa := \na_{\al} E_\fa^{\al} = 0 \, . 
\eeq
Finally, we also have the primary constraint from \eqref{eq:Edef} 
\beq \label{eq:preHal}
\cD'_{\al} := \vep_{\al\be\ga} E^{\be}_{\fa} B^{\fa\ga} = 0 \, ,
\eeq 
which we will refer to as the ``Poynting'' constraint, since it is the generalization of the Poynting vector to the non-abelian group case. Once combined with the Gauss constraint, it leads to the usual ``diffeomorphism'' constraint associated with 3-diffeomorphisms on $\Si_t$
\beq
\cD_{\al} := \cD'_{\al} - A_{\al}^\fa \cG_\fa \equiv E_\fa^{\be} \( \pa_{\al} A_{\be}^\fa - \pa_{\be} A_{\al}^\fa \) - A_{\al}^\fa \pa_{\be} E_\fa^{\be} = 0 \, .
\eeq
Thus, the {\it generic} canonical action reads
\beq \label{eq:Scangen}
S = \int \ed^4 x \[ E_\fa^{\al} \dot{A}_{\al}^\fa - \ti{\ph} \cH(\bm{\psi}) - N^{\al} \cD_{\al} - \te^\fa \cG_\fa \] \, ,
\eeq
where $N^{\al}$ is introduced in order to impose the diffeomorphism constraint. We thus see that $\cH(\bm{\psi}(E,B))$ is the (de-densitized) Hamiltonian constraint, while $\ti{\ph}$ plays the role of a densitized lapse function. Given \eqref{eq:Edef}, the shift \eqref{eq:Xishift} discussed in the previous section amounts to a redefinition of the conjugate momenta
\beq
E_{\fa}^{\al} \to E_{\fa}^{\al} + z_{\fa\fb} B^{\fb\al} \, . 
\eeq
In the action \eqref{eq:Scangen} the corresponding variation of $E_\fa^{\al} \dot{A}_{\al}^\fa$ is the total time-derivative of the Chern-Simons 3-form of $A_{\al}^{\fa}$, so this is a canonical transformation. Moreover, $\cD'_{\al}$ and $\cG_{\fa}$ are invariant, so we simply obtain a different Hamiltonian constraint leading to equivalent classical physics, as was already noticed at the Lagrangian level.

As one could expect, in the generic case \eqref{eq:Scangen} contains the minimal amount of constraints that are implied by the local symmetries of the theory. Because of this, they must form a first-class system under the Poisson bracket 
\beq \label{eq:Poisson}
\{ \Ord, \Ord' \} := \int \ed^3 x \[ \frac{\de \Ord}{\de A_{\al}^{\fa}} \frac{\de \Ord'}{\de E^{\al}_{\fa}} - \frac{\de \Ord'}{\de A_{\al}^{\fa}} \frac{\de \Ord}{\de E^{\al}_{\fa}} \] \, .
\eeq
Thus, for a gauge group of dimension $D$, the canonical action \eqref{eq:Scangen} leads to the following DoF count: $6D$ canonical fields, minus $4 + D$ combinations that are neutralized by the constraints, minus $4 + D$ that are neutralized by the associated gauge transformations equal $4 \( D - 2 \)$ reduced phase space variables. In the case of SO(1,3) we have $D = 6$ and therefore $16$ reduced phase space variables, or eight propagating DoF, showing that the generic theory of the form \eqref{eq:Sf} has indeed the spectrum of a bigravity theory, as discussed in the Introduction. This described DoF count corresponds to the generic case, because we already know of at least one case (GR) where extra constraints must appear in order to reduce the number of DoF down to two. 

On the other hand, when $G={\rm SO}(3)$ or ${\rm SO}(1,2)$ we have $D = 3$, and so we find $2$ propagating DoF, just as in GR. This is also the simplest case, studied in details in the series of works \cite{Capovilla:1989ac, Bengtsson:1992cm, Bengtsson:1992dq} and \cite{Krasnov:2008fm,Krasnov:2009ik, Krasnov:2011up, Krasnov:2011pp,Celada:2015jda}. We have
\beq
\fa \to i \in \{ 1,2,3 \} \hspace{1cm} \ka_{ij} = {\rm diag}(\pm1,1,1) \, , \hspace{1cm} f^i_{\,\,\,jk} = \ka^{il} \vep_{ljk} \, ,
\eeq
and the relation \eqref{eq:Edef} completely determines $\psi_{ij}$
\beq \label{eq:XiofEB3}
\psi^{ij} := E^{(i}_{\al} B^{j)\al} \, ,
\eeq
where $E_{\al}^i$ is the inverse matrix of $E_i^{\al}$ and one has to take into account the Poynting constraint \eqref{eq:preHal} to satisfy \eqref{eq:Edef}. Euclidean GR can be described in this formalism and corresponds to 
\beq  \label{eq:HGRsd}
\cH_{\rm GR}(\bm{\psi}) = \La - M^2\, \Tr \( \bm{\psi} \) \, ,
\eeq 
and with \eqref{eq:XiofEB3} we have
\beq
\cH_{\rm GR} = \La - M^2 E_{\al}^i B_i^{\al} \, .
\eeq
One recognizes the (de-densitized) Hamiltonian constraint of Ashtekar Hamiltonian formulation \cite{Ashtekar:1987gu} of GR.

\section{Vacuum solution and linearised theory} \label{sec:vaclin}

We now specialise to the case of Lorentz group $G={\rm SO}(1,3)$. The condition \eqref{eq:Ldeg1} implies that $f$ is not polynomial in $\bm{X}$, except for the topological case $f \( \bm{X} \) = \Tr \[ \bm{z} \bm{X} \]$. As a result, the action functional will not be smooth around the configuration $\bm{X} = 0$, meaning that one cannot use the latter as a vacuum for perturbative computations. In the case of GR, $\bm{X} = 0$ corresponds to flat space-time and the unavailability of that configuration can directly be seen in the fact that the cosmological constant appears in the denominator in  \eqref{eq:SGR}. One does have access, however, to the other ``maximally symmetric'' configurations $\bar{A}^{ab}$, i.e. the case where there exist vierbein 1-forms $\bar{e}^a$ and a constant $C \neq 0$ with curvature dimensions such that\footnote{Observe that only the first equation is actually needed, as the second one is obtained by acting with the exterior covariant derivative and using the Bianchi identity $\ced F^{ab} \equiv 0$, along with some simple algebraic manipulations. \label{ft:AdS}}
\beq \label{eq:MSsol}
\bar{F}^{ab} = C \bar{e}^a \we \bar{e}^b \, , \hspace{1cm} \bar{\ced} \bar{e}^a := \ed \bar{e}^a + \bar{A}^a_{\,\,\,b} \we \bar{e}^b = 0 \, .
\eeq
To see that this $\bar{A}^{ab}$ is a solution of all theories \eqref{eq:Sf} for all $C$, we note that 
\beq
\bar{X}^{abcd} \equiv - \frac{1}{2}\, C^2 \vep^{abcd} \bar{e} \, , \hspace{1cm} \bar{e} := \frac{1}{4!}\, \vep_{abcd}\, \bar{e}^a \we \bar{e}^b \we \bar{e}^c \we \bar{e}^d \, ,
\eeq
or, in matrix notation,
\beq \label{eq:Xbar}
\bar{\bm{X}} \equiv - \frac{1}{2}\, C^2 \bar{e} \, \star \, ,
\eeq
so $\bar{\bm{\psi}} := \bm{\psi}(\bar{\bm{X}})$ is a constant invariant matrix and therefore \eqref{eq:EOM} is satisfied. Thus, although the action \eqref{eq:Sf} only involves dimensionless parameters, the requirement $\bm{X} \neq 0$ necessarily reintroduces a reference scale through the solutions. Moreover, \eqref{eq:MSsol} also reintroduces a notion of space-time, through the (A)dS vierbein $\bar{e}^a$, which will then serve as a background geometry for the fluctuations. These are defined by 
\beq
\cA^{ab} := A^{ab} - \bar{A}^{ab} \, , \hspace{1cm} F^{ab} \equiv \bar{F}^{ab} + \cF^{ab} + \cA^a_{\,\,\,c} \we \cA^{cb} \, , \hspace{1cm} \cF^{ab} := \bar{\ced} \cA^{ab} \, ,
\eeq
and we are interested in the part of the Lagrangian that is second order in $\cA^{ab}$, i.e. the linearised theory. We find
\beq \label{eq:L2}
f_{(2)} := \frac{1}{2} \left. \de^2 f \right|_{A = \bar{A}} \equiv \frac{1}{2} \[ \bar{H}_{\fa\fb,\fc\fd}\, \de X^{\fa\fb} \de X^{\fc\fd} + \bar{G}_{\fa\fb} \de^2 X^{\fa\fb} \] \, ,
\eeq
where the Hessian $H_{\fa\fb,\fc\fd}$ is defined in \eqref{eq:LabcdXcd0}. Note that the second term in this expression is the second variation of topological terms
\beq
\bar{G}_{\fa\fb} X^{\fa\fb} \propto \al\, \vep_{abcd} F^{ab} \we F^{cd} + \be F_{ab} \we F^{ab} \, ,
\eeq
so we can ignore it, meaning that we only need the first variation
\beq
\de X^{\fa\fb} \equiv \bar{F}^{(\fa} \we \cF^{\fb)} \, , \hspace{1cm} \de X^{abcd} \equiv \frac{1}{2}\, C \[ \bar{e}^a \we \bar{e}^b \we \cF^{cd} + \bar{e}^c \we \bar{e}^d \we \cF^{ab} \] \, .
\eeq
As for $\bar{H}_{\fa\fb,\fc\fd}$, we note that \eqref{eq:Xbar} makes it a combination of tensor products of $\bm{1}$ and $\star$ which must, however, satisfy the background version of \eqref{eq:LabcdXcd0}. Given \eqref{eq:Xbar}, that is $\bar{H}_{\fa\fb,\fc\fd}\, \star^{\fc\fd} = 0$, which leaves us with the four-parameter expression
\bea
\bar{H}_{\fa\fb,\fc\fd} & \equiv & C^{-2} \bar{e}^{-1} \[ c_1 \star_{\fa(\fc} \star_{\fd)\fb} + \frac{1}{6}\, c_2 \star_{\fa\fb} \star_{\fc\fd} + \( c_1 + c_2 \) \bm{1}_{\fa(\fc} \bm{1}_{\fd)\fb} + \frac{1}{3}\, c_3 \bm{1}_{\fa\fb} \bm{1}_{\fc\fd} \right. \nn \\
 & & \left.  \hspace{2cm} +\, \ti{c} \( \star_{\fa(\fc} \bm{1}_{\fd)\fb} + \bm{1}_{\fa(\fc} \star_{\fd)\fb} - \frac{1}{3} \( \star_{\fa\fb} \bm{1}_{\fc\fd} + \bm{1}_{\fa\fb} \star_{\fc\fd} \) \) \] \, . \label{eq:Hofcs}
\eea
It will be more convenient to express the final result using only indices of a given kind, and here we will choose for definiteness diffeomorphism indices. We thus define
\beq
\cA_{\mu\nu\ro} := \bar{e}_{\mu}^a \bar{e}_{\nu}^b \cA_{ab\ro} \, , \hspace{1cm} \cF_{\mu\nu\ro\si} := \bar{e}_{\mu}^a \bar{e}_{\nu}^b \cF_{ab\ro\si} \equiv \bar{\na}_{\ro} \cA_{\mu\nu\si} - \bar{\na}_{\si} \cA_{\mu\nu\ro} \, ,
\eeq
where $\bar{\na}$ denotes the torsion-free covariant derivative compatible with the background metric  
\beq
\bar{g}_{\mu\nu} := \et_{ab} \bar{e}^a_{\mu} \bar{e}^b_{\nu} \, .
\eeq
Using the latter to displace the corresponding indices, we will also need the traces of the field strength
\beq
\cF_{\mu\nu} := \cF^{\ro}_{\,\,\,\mu\ro\nu} \, , \hspace{1cm} \cF := \cF_{\mu}^{\mu} \, ,
\eeq
its traceless part 
\beq
\cC_{\mu\nu\ro\si} := \cF_{\mu\nu\ro\si} - \frac{1}{2} \( \bar{g}_{\mu\ro} \cF_{\nu\si} - \bar{g}_{\mu\si} \cF_{\nu\ro} - \bar{g}_{\nu\ro} \cF_{\mu\si} + \bar{g}_{\nu\si} \cF_{\mu\ro} \) + \frac{1}{6} \( \bar{g}_{\mu\ro} \bar{g}_{\nu\si} - \bar{g}_{\mu\si} \bar{g}_{\nu\ro} \) \cF \, , 
\eeq
the irreducible part thereof
\beq
\cW_{\mu\nu\ro\si} :=\frac{1}{3} \[ \cC_{\mu\nu\ro\si} + \frac{1}{2} \( \cC_{\mu\ro\nu\si} - \cC_{\mu\si\nu\ro} - \cC_{\nu\ro\mu\si} + \cC_{\nu\si\mu\ro} \) + \cC_{\ro\si\mu\nu} \] \, , 
\eeq
which has the symmetries of the Weyl tensor, and
\beq
\ti{\cC}_{\mu\nu\ro\si} := \frac{1}{2}\, \bar{\ep}_{\mu\nu}^{\,\,\,\,\,\,\,\ka\la} \cC_{\ka\la\ro\si} \, ,  \hspace{1cm} \bar{\ep}_{\mu\nu\ro\si} := \sqrt{-\bar{g}}\, \vep_{\mu\nu\ro\si} \, ,
\eeq
which has the symmetries of the Riemann tensor \cite{Mitsou:2019nlt}. In terms of these quantities, using \eqref{eq:Hofcs}, we finally obtain the nontrivial part of \eqref{eq:L2} 
\beq \label{eq:L2gen}
f_{(2)} = \frac{1}{8}\, \bar{e} \[ \( c_1 - c_3 \) \cC_{\mu\nu\ro\si} \cC^{\ro\si\mu\nu} - \( c_1 + c_2 + c_3 \) \cC_{\mu\nu\ro\si} \cC^{\mu\nu\ro\si} + 2 c_3 \cW_{\mu\nu\ro\si} \cW^{\mu\nu\ro\si} + 2 \ti{c} \cC_{\mu\nu\ro\si} \ti{\cC}^{\mu\nu\ro\si} \] \, .
\eeq
Thus, all pure spin connection theories depend exclusively on the traceless component $\cC_{\mu\nu\ro\si}$ of the field strength at the linear level. Note that this is contrary to what was conjectured in \cite{Mitsou:2019nlt} as being a unique property of GR. Rather, the case of GR is the restriction from four to two independent parameters $M^2/C$ and $\be$, i.e. including the overall normalization, \cite{Zinoviev:2005qp,Basile:2015jjd,Mitsou:2019nlt}
\beq \label{eq:clinGR}
c^{\rm GR}_1 = - \frac{\( 2 + \be^2 \)}{4} \frac{M^2}{C} \, , \hspace{1cm} c^{\rm GR}_2 = \frac{1}{2} \frac{M^2}{C} \, , \hspace{1cm} c^{\rm GR}_3 = \frac{\be^2}{4} \frac{M^2}{C} \, , \hspace{1cm} \ti{c}^{\rm GR} = \frac{\be}{2} \frac{M^2}{C} \, ,
\eeq
with $C = \La/3$. There is clearly an infinite-parametric set of defining functions $f$ that reproduce the linearised action of GR, since this requirement only constrains the second-derivative of $f$, the Hessian matrix, evaluated on the (A)dS background. To obtain the full spectrum of these theories, one must therefore perform a canonical analysis at the fully non-linear level.

\section{Canonical analysis and GR uniqueness} \label{sec:GRuniq}

We now use the tools developed in section \ref{sec:genprop} to perform the canonical analysis of the theories \eqref{eq:Sf} with the gauge group SO$(1,3)$. In particular, we wish to identify the cases where extra constraints emerge, thus leading to potentially viable theories that are not GR. We will use the algebra indexation $\fa,\fb,\fc,\dots$ introduced previously.

\subsection{Hamiltonian in terms of phase space variables} \label{sec:extraprimconstr}

Contrary to the case $G = {\rm SO}(3)$ considered at the end of subsection \ref{sec:caness}, here the momentum/velocity relation \eqref{eq:Edef} does not fully determine $\bm{\psi}$ in terms of the canonical variables, because the group dimension is larger than three. More precisely, $\bm{\psi}$ is an invertible $6 \times 6$ matrix, so its effect on 3 vectors determines half of its rows, but it is also symmetric, so this also determines half of its columns. The undetermined information in \eqref{eq:Edef} therefore amounts to a symmetric $3 \times 3$ matrix. That information is determined by the following identity in configuration space (remember that $X_{\fa\fb}$ is the inverse of $X^{\fa\fb}$)
\beq  \label{eq:speciden6}
X_{\fa\fb} B^{\fa\al} B^{\fb\be} \equiv 0 \, ,
\eeq
which, through the inversion \eqref{eq:XofLgen}, leads to the following symmetric $3 \times 3$ matrix equation for $\bm{\psi}$
\beq  \label{eq:speciden7}
\hat{X}_{\fa\fb}(\bm{\psi}) \, B^{\fa\al} B^{\fb\be} = 0 \, .
\eeq
To show \eqref{eq:speciden6}, write it as $\propto \vep_{\fa_1 \dots \fa_6} \vep_{\fb_1 \dots \fb_6} X^{\fa_1 \fb_1} \dots X^{\fa_5 \fb_5} B^{\fa_6\al} B^{\fb_6\be}$, use \eqref{eq:XofKB} and then observe that each term in the corresponding sum contains the antisymmetrization of at least four $B^{\fa\al}$ or four $K_{\al}^{\fa}$ fields, which yields zero since $\al$ is a 3-dimensional index. Thus, in the generic case equations \eqref{eq:Edef} and \eqref{eq:speciden7} completely determine $\bm{\psi}$ in terms of the electric and magnetic fields, meaning the same for the Hamiltonian constraint $\cH(\bm{\psi})$ in \eqref{eq:Scangen}. To construct an explicit expression for $\bm{\psi}$, we define the metric density $E_{\al\be}$ as the inverse matrix of
\beq 
E^{\al\be} := \ka^{\fa\fb} E_{\fa}^{\al} E_{\fb}^{\be} \, .
\eeq
We can then use this to define the electric fields with reversed index positions
\beq
E_{\al}^{\fa} := \ka^{\fa\fb} E_{\al\be} E_{\fb}^{\be} \, ,
\eeq 
which behaves as an ``inverse'' of $E_{\fa}^{\al}$ when contracting the algebra indices
\beq
E_{\fa}^{\al} E_{\be}^{\fa} \equiv \de^{\al}_{\be} \, . 
\eeq
However, since this is a $3 \times 6$ matrix, this relation does not hold when contracting the space indices
\beq
E_{\fa}^{\fb} := E_{\fa}^{\al} E_{\al}^{\fb} \, , 
\eeq
and we find instead the projector onto the subspace spanned by the three $E_{\fa}^{\al}$
\beq \label{eq:EEEBBB}
E_{\fa}^{\fc} E_{\fc}^{\fb} \equiv E_{\fa}^{\fb} \, , \hspace{1cm} E_{\fa}^{\fb} E_{\fb}^{\al} \equiv E_{\fa}^{\al} \, .
\eeq
One can then verify that
\beq \label{eq:Labexpl}
\psi^{\fa\fb} = E_{\al}^{(\fa} \[ 2 \de^{\fb)}_{\fc} - E^{\fb)}_{\fc} \] B^{\fc\al}  + M^{\fa\fb} \, ,  \hspace{1cm} M^{\fa\fb} E_{\fb}^{\al} \equiv 0 \, ,
\eeq
satisfies the desired relation $B^{\fa\al} = \psi^{\fa\fb} E_{\fb}^{\al}$ (see \eqref{eq:Edef}) if one uses the Poynting constraint \eqref{eq:preHal}. Here the matrix $\bm{M}$ parametrizes the information that is left undetermined by \eqref{eq:Edef}. This is a $6 \times 6$ symmetric matrix that is normal to three independent vectors, so it carries the information of a symmetric $3 \times 3$ matrix indeed. Thus, if we insert \eqref{eq:Labexpl} inside \eqref{eq:speciden7}, the matrix $\bm{M}$ is generically fully determined, so is the desired relation $\bm{\psi} = \bm{\psi}(E,B)$ and ultimately the Hamiltonian constraint $\cH = \cH(\bm{\psi}(E,B))$.

\subsection{Extra primary constraints}

We are now in a position to identify the set of ``special'' cases, i.e. the theories for which extra constraints appear. As we already discussed, GR must be one of such theories. These are the theories for which \eqref{eq:speciden7} does not fully determine $\bm{M}$, thus turning part of the equations \eqref{eq:speciden7} into phase space constraints. There are several possibilities for partially determining $\bm{M}$, because the tensor $M^{abcd}$ is reducible under the local SO(1,3) symmetry. As we will see, the case of GR corresponds to a completely undetermined $\bm{M}$. This leads to six extra primary constraints, whose conservation then implies another six secondary constraints, thus reducing the 16 canonical DoF of the generic theory down to 4. In this paper we are primarily interested in theories with the same DoF count as GR, so we look for theories for which all of $\bm{M}$ is undetermined.

We therefore look for matrix functions $\hat{\bm{X}}(\bm{\psi})$ for which \eqref{eq:speciden7} is completely independent of $\bm{M}$. Such matrix functions generalise what one has in the case of GR \eqref{eq:XofLGR}. Indeed, consider
\beq \label{eq:magicf}
\hat{\bm{X}}^{-1}(\bm{\psi}) = \bm{\psi}^{-1} \bm{z}_0^{-2} \bm{\psi}^{-1} + \bm{z}_1 \bm{\psi}^{-1} + \bm{\psi}^{-1} \bm{z}_1 - \bm{z}_2^2 \, ,
\eeq
where $\bm{z}_{0,1,2}$ are constant invariant matrices \eqref{eq:zalbe}. Inserting this expression in \eqref{eq:speciden7}, and then using $B^{\fa\al} = \psi^{\fa\fb} E_{\fb}^{\al}$ and \eqref{eq:Labexpl}, we observe that all $\bm{M}$-dependent terms vanish, because they contain at least one $M^{\fa\fb} E_{\fb}^{\al}$ contraction. What is left is the primary constraints
\beq \label{eq:extraconstr}
\[ \bm{z}_0^{-2} \]^{\fa\fb} E_{\fa}^{\al} E_{\fb}^{\be} + 2 [\bm{z}_1]^{\fa}_{\fb} E_{\fa}^{(\al} B^{\fb\be)} - \[ \bm{z}_2^2 \]_{\fa\fb} B^{\fa\al} B^{\fb\be} = 0 \, .
\eeq
Moreover, any extension of \eqref{eq:magicf} will necessarily introduce $\bm{M}$ terms that are not contracted with electric fields and thus remain, so \eqref{eq:magicf} is the most general matrix function that leads to six primary constraints. For this special subset of theories, one can simply set $\bm{M} = 0$ in \eqref{eq:Labexpl}. Indeed, different choices of $\bm{M}$ matrix lead to different $\bm{\psi}(E,B)$ relations, but they all satisfy both required equations \eqref{eq:Edef} and \eqref{eq:speciden7} on the constraint surface. Thus, here too the $\bm{\psi}$ matrix and the Hamiltonian $\cH(\bm{\psi})$ are determined in terms of $E$ and $B$. Finally, as we will see, the parameters in $\bm{z}_{0,1,2}$ are not all independent, but must satisfy some ``integrability conditions'', since the relation \eqref{eq:magicf} must derive from a Hamiltonian function \eqref{eq:hatXofH}.

Comparing \eqref{eq:magicf} with \eqref{eq:XofLGR}, and keeping in mind that the normalization of $\hat{\bm{X}}$ is irrelevant since it can be reabsorbed inside $\ph$, we recognize the case of GR when $\bm{z}_0$ is given by \eqref{eq:LGR}, and $\bm{z}_1 = \bm{z}_2 = 0$. The corresponding extra constraints \eqref{eq:extraconstr} are the simplicity constraints of the covariant canonical formulation of vierbein GR \cite{Alexandrov:2000jw, Alexandrov:2008fs}. They reduce to the more familiar $\vep^{abcd} E_{ab}^{\al} E_{cd}^{\be} = 0$ in the parity-even case $\be = 0$. These constraints can then be solved $E_{ab}^{\al} \propto \vep_{abcd} \, \vep^{\al\be\ga} e^c_{\be} e^d_{\ga}$, which is how the full vierbein information $\{ \ti{\ph}, N^{\al}, e_{\al}^a \}$ reappears in this approach. 

Now remember that shifting $\bm{\psi}^{-1}$ by a constant matrix \eqref{eq:Xishift} amounts to adding a topological term in the action \eqref{eq:fshift}, meaning that not all of the different $\bm{z}_{0,1,2}$ choices lead to different classical theories. We can therefore use this freedom to reduce the special cases \eqref{eq:magicf} as follows.

\subsubsection{The ``deformed GR'' theory}

First, in the case where the theory is a deformation of GR, i.e. $\bm{z}_0^{-2} \neq 0$, we can perform the shift
\beq
\bm{\psi}^{-1} \to \bm{\psi}^{-1} - \bm{z}_0^{-2} \bm{z}_1 \, ,
\eeq 
to make \eqref{eq:magicf} of the form (after redefining $\bm{z}_2$)
\beq \label{eq:deftGRXi}
\hat{\bm{X}}^{-1}(\bm{\psi}) = \bm{\psi}^{-1} \bm{z}_0^{-2} \bm{\psi}^{-1} - \bm{z}_2^2  \, ,
\eeq
and the corresponding constraint \eqref{eq:extraconstr} is
\beq \label{eq:extratGRconstr}
\[ \bm{z}_0^{-2} \]^{\fa\fb} E_{\fa}^{\al} E_{\fb}^{\be} - \[ \bm{z}_2^2 \]_{\fa\fb} B^{\fa\al} B^{\fb\be} = 0 \, .
\eeq  
Next, the integrability condition \eqref{eq:hatXofH} forces the $\bm{z}_{0,2}$ matrices to be proportional to each other $\bm{z}_2 = \al \bm{z}_0 \equiv \al \bm{z}$, in which case the constraint reads
\beq \label{eq:deformedGRconstr}
\cC^{\al\be} := \[ \bm{z}^{-2} \]^{\fa\fb} E_{\fa}^{\al} E_{\fb}^{\be} - \al^2 \[ \bm{z}^2 \]_{\fa\fb} B^{\fa\al} B^{\fb\be} = 0 \, .
\eeq  
Indeed, only in that case can we express \eqref{eq:deftGRXi} as an algebraic relation between only two matrices (apart from the identity)
\beq
\hat{\bm{X}}_{\bm{z}}^{-1} = \bm{\psi}_{\bm{z}}^{-2} - \al^2 \bm{1} \, ,  \hspace{1cm} \bm{Y}_{\bm{z}} := \bm{z} \bm{Y} \bm{z} \, ,
\eeq
which can therefore be integrated without the obstruction of matrix non-commutativity. We find 
\beq \label{eq:deformedGRH}
\cH(\bm{\psi}) = \La - \frac{M^2}{\al} \, \Tr \, {\rm arctanh} \( \al \bm{\psi}_{\bm{z}} \) \, ,
\eeq
and one can readily verify that we consistently recover the GR result \eqref{eq:HGR} in the $\al \to 0$ limit. This theory is therefore a 1-parameter deformation of GR, so we will refer to it as the ``deformed GR theory''. To obtain the corresponding $f(F \we F)$ form, we insert \eqref{eq:deformedGRH} inside \eqref{eq:SH} and integrate out $\bm{\psi}$. The solution to its equation of motion is 
\beq
\bm{\psi}_{\bm{z}} = \pm \, \sqrt{\frac{\bm{X}_{\bm{z}}}{M^2 \ph \bm{1} + \al^2 \bm{X}_{\bm{z}}}} \, ,
\eeq
so the resulting action reads
\beq \label{eq:Sdefint}
S = \int \[ \pm\, \Tr \sqrt{\bm{X}_{\bm{z}} \( M^2 \ph \bm{1} + \al^2 \bm{X}_{\bm{z}} \)} - \ph \( \La \mp \frac{M^2}{\al} \, \Tr \, {\rm arctanh} \,\sqrt{\frac{\al^2 \bm{X}_{\bm{z}}}{M^2 \ph \bm{1} + \al^2 \bm{X}_{\bm{z}}}} \) \] \, .
\eeq
We must now further integrate out $\ph$, but its equation of motion
\beq \label{eq:phofXzdef}
\pm \Tr \, {\rm arctanh} \, \sqrt{\frac{\al^2 \bm{X}_{\bm{z}}}{M^2 \ph \bm{1} + \al^2 \bm{X}_{\bm{z}}}} = \frac{\al\La}{M^2} \, ,
\eeq
is not easy to solve. Nevertheless, this equation simplifies the action (it is the Hamiltonian constraint), so one compact way of describing the theory is
\beq 
S = \pm \int \Tr \sqrt{\bm{X}_{\bm{z}} \( M^2 \ph(\bm{X}_{\bm{z}})\, \bm{1} + \al^2 \bm{X}_{\bm{z}} \)} \, ,
\eeq
with $\ph(\bm{X}_{\bm{z}})$ given implicitly by \eqref{eq:phofXzdef}. Alternatively, we can expand in powers of $\al$ and solve for $\ph$ perturbatively, thus obtaining a peek at the theory close to GR
\beq \label{eq:defGRS}
S = \frac{M^2}{\La} \int \[ \Tr \sqrt{\bm{X}_{\bm{z}}} \]^2 \[ 1 + \frac{\al^2\La^2}{3M^4} \frac{\Tr \bm{X}_{\bm{z}}^{3/2}}{\[ \Tr \bm{X}_{\bm{z}}^{1/2} \]^3} + \Ord(\al^4) \] \, .
\eeq
However, this is not very useful in practice, because $\bm{X}_{\bm{z}}$ has a non-zero VEV, so that the higher orders in $\al$ cannot be neglected. Finally, note that the opposite limit $\al \to \infty$ is singular, a fact which will be relevant later.

\subsubsection{The ``determinant'' theory}

Now if $\bm{z}_0^{-2} = 0$, but $\bm{z}_1 \neq 0$, we can shift $\bm{\psi}^{-1}$ as follows 
\beq
\bm{\psi}^{-1} \to \bm{\psi}^{-1} + \frac{1}{2}\, \bm{z}_1^{-1} \bm{z}_2^2 \, ,
\eeq 
to obtain
\beq \label{eq:defnGRXi}
\hat{\bm{X}}^{-1}(\bm{\psi}) = \bm{z}_1 \bm{\psi}^{-1} + \bm{\psi}^{-1} \bm{z}_1 \, ,
\eeq
and therefore the constraint \eqref{eq:extraconstr}
\beq \label{eq:extranGRconstr}
2 [\bm{z}_1]^{\fa}_{\fb} E_{\fa}^{(\al} B^{\fb\be)} = 0 \, .
\eeq 
The integrability condition \eqref{eq:hatXofH} leads to $\bm{z}_1 = \al \bm{1}$, so that the extra constraint of this theory is
\beq \label{eq:determinantconstr}
\cC^{\al\be} := 2 E_{\fa}^{(\al} B^{\fa\be)} = 0 \, ,
\eeq 
and its Hamiltonian constraint is
\beq \label{eq:determinantH}
\cH(\bm{\psi}) = \la - \frac{1}{2\al} \,\Tr \log \bm{\psi} \, .
\eeq
Proceeding as in the previous subsection for the corresponding pure-connection formulation, this time there is no obstruction to integrating out both $\bm{\psi}$ and $\ph$ and we find
\beq \label{eq:detS}
S = 6 e^{-\al\la/3} \int \[ \det \bm{X} \]^{1/6}  \, ,
\eeq
so we dub this the ``determinant'' theory. Contrary to deformed GR, note that this theory is disconnected from GR in parameter space and has two interesting properties. First, the constraint \eqref{eq:extranGRconstr} combines with the Poynting one \eqref{eq:preHal} to make the electric and magnetic fields orthogonal in their internal indices 
\beq \label{eq:EBnormal}
E_{\fa}^{\al} B^{\fa\be} = 0 \, .
\eeq
Thanks to this, \eqref{eq:Labexpl} simplifies to
\beq \label{eq:psidettheory}
\psi^{\fa\fb} = 2 E_{\al}^{(\fa} B^{\fb)\al} \, ,
\eeq
and (see appendix \ref{app:detpsi})
\beq
\Tr \log \bm{\psi} \equiv \log \det \bm{\psi} = \log \( - E^{-1} B \) \, , 
\eeq
where
\beq
E := \det E^{\al\be} \, , \hspace{1cm} B := \det B^{\al\be} \, , \hspace{1cm} B^{\al\be} := \ka_{\fa\fb} B^{\fa\al} B^{\fb\be} \, . 
\eeq
Thus, the Hamiltonian constraint is simply the proportionality of the determinants of the two 3-metric densities $E = - e^{2\al\la} B$. The second interesting property is that the linearised theory around (A)dS, found using \eqref{eq:detS} and the procedure outlined in section \ref{sec:vaclin}, is the same as GR with vanishing Immirzi parameter $\be = 0$. This is quite remarkable given the fact that the action of this theory is clearly not continuously connected to the one of GR.

\subsubsection{Self-Dual Gravity} 

The last independent case of \eqref{eq:magicf} is the one with $\bm{z}_0^{-2} = \bm{z}_1 = 0$, i.e.
\beq \label{eq:XofpsiSDG}
\hat{X}(\bm{\psi}) = - \bm{z}_2^{-2} \, ,
\eeq
so the corresponding constraints appear as the ``dual'' of the ones of GR
\beq \label{eq:CSDG}
\[ \bm{z}_2^2 \]_{\fa\fb} B^{\fa\al} B^{\fb\be} = 0 \, ,
\eeq
and the Hamiltonian function is
\beq \label{eq:HGRdual}
\cH(\bm{\psi}) = \La - M^2 \Tr \( \bm{z}_2^{-1} \bm{\psi}^{-1} \bm{z}_2^{-1} \) \, .
\eeq
This looks similar to the GR case \eqref{eq:HGR}, only $\bm{\psi}$ is replaced by $\bm{\psi}^{-1}$, which then allows one to get rid of the cosmological constant term $\sim \La$ through the shift freedom \eqref{eq:Xishift}. From the form of the action \eqref{eq:SH} we then see that $\psi_{\fa\fb}$ enters linearly, so that it cannot be integrated out to reach a pure connection formulation, but instead imposes the constraint 
\beq \label{eq:HGRdualconstrto}
\frac{1}{2}\, F^{\fa} \we F^{\fb} = - \ph \[ \bm{z}_2^{-2} \]^{\fa\fb} \, .
\eeq 
This theory was first studied in \cite{Krasnov:2016emc} for the 3-dimensional groups and is known as ``Self-Dual Gravity'', because of its analogy to self-dual Yang-Mills theory. Because of \eqref{eq:XofpsiSDG}, for this theory to have the (A)dS configuration as a solution \eqref{eq:Xbar}, we must have $\bm{z}_2^2 \propto \star$ and thus $\cH(\bm{\psi}) \propto \Tr \[ \star\, \bm{\psi}^{-1} \]$. The reader familiar with the real Plebanski (or ``$BF$'') formulation of GR \cite{Plebanski:1977zz,DePietri:1998hnx} will then immediately notice that \eqref{eq:HGRdualconstrto} takes the form of the simplicity constraint for the $B$ fields, meaning that we have one of the following two options
\beq \label{eq:GRdual2opt}
F^{ab} = C e^a \we e^b \, , \hspace{1cm} {\rm or} \hspace{1cm} F^{ab} = \frac{1}{2}\, C \vep^{abcd} \, e_c \we e_d \, ,
\eeq
for a set of 1-forms $e^a$, where $C$ is a constant and $\phi \propto \vep_{abcd}\, e^a \we e^b \we e^c \we e^d$. As already noted in footnote \ref{ft:AdS}, taking the exterior derivative of \eqref{eq:GRdual2opt} we find, after some algebraic manipulations, $\ced e^a = 0$, meaning that $A^{ab}$ is the torsion-free spin connection of the vierbein $e^a$. On the other hand, wedging \eqref{eq:GRdual2opt} with $e^b$ and using $0 = \ced^2 e^a \equiv F^a_{\,\,\,b} \we e^b$, we obtain zero on both sides of the first equation, but not for the second equation, meaning that only the first option is possible. In conclusion, \eqref{eq:HGRdualconstrto} implies that this theory admits only the (A)dS solution for $A^{ab}$, so there are no DoF in that field. There are DoF in $\psi_{\fa\fb}$, however, since this field satisfies a dynamical equation \eqref{eq:EOM}. Nevertheless, the corresponding excitations cannot have an energy that is bounded from below, because $\psi_{\fa\fb}$ enters the action linearly, so this theory is not physical.

Finally, note that the constraint \eqref{eq:CSDG} appears as the $\al \to \infty$ limit of the one of deformed GR \eqref{eq:deformedGRconstr}. This explains why the pure-connection action of deformed GR is singular in that limit, since $\bm{\psi}$ can no longer be integrated out. Moreover, it shows that the $\al$ parameter of deformed GR continuously connects GR and Self-Dual Gravity.

\subsection{Dirac algorithm}
    
We thus identified two potentially interesting theories with extra primary constraints in the canonical action \eqref{eq:Scangen}, i.e. deformed GR \eqref{eq:deformedGRconstr}, \eqref{eq:deformedGRH} and the determinant theory \eqref{eq:determinantconstr}, \eqref{eq:determinantH}. The canonical action of these theories is then \eqref{eq:Scangen} supplemented by a term imposing the extra constraints through a Lagrange multiplier $\la_{\al\be}$  
\beq \label{eq:ScanC}
S = \int \ed^4 x \[ E_\fa^{\al} \dot{A}_{\al}^\fa - \ti{\ph} \cH - N^{\al} \cD_{\al} - \te^\fa \cG_\fa - \la_{\al\be} \cC^{\al\be} \] \, .
\eeq
We must now ensure that the constraint surface is conserved under the time-evolution, i.e. apply Dirac's algorithm for constrained Hamiltonian systems. In this procedure we can neglect the diffeomorphism and Gauss constraints $\cD_{\al}$ and $\cG_{\fa}$, respectively, since they commute with all other constraints on-shell, thanks to the explicit covariance under the gauge symmetries they respectively generate. We can therefore focus on the commutation relations of $\cH$ and $\cC^{\al\be}$, computed through the smeared constraints  
\beq
H[\ti{\ph}] := \int \ed^3 x\, \ti{\ph} \cH \, , \hspace{1cm} C[\la] := \int \ed^3 x\, \la_{\al\be} \cC^{\al\be} \, .
\eeq
The conservation equation of $\cC^{\al\be}$ reads 
\beq \label{eq:dotC}
\dot{\cC}^{\al\be} = \ti{\ph} \, \cC'^{\al\be} + W^{\al\be,\ga\de} \la_{\ga\de} = 0 \, , 
\eeq
where, schematically (up to Dirac delta factors),
\beq
\cC'^{\al\be} := \{ \cH, \cC^{\al\be} \} \, ,  \hspace{1cm} W^{\al\be,\ga\de} := \{ \cC^{\al\be}, \cC^{\ga\de} \} \, .
\eeq
As for the one of $\cH$, we have
\beq \label{eq:dotH}
\dot{\cH} = - \, \la_{\al\be} \cC'^{\al\be} = 0 \, ,
\eeq
where have used the fact that, for all $f(F\we F)$ theories,
\beq \label{eq:HHcom}
\{ H[\ti{\ph}], H[\ti{\ph}'] \} = \int \ed^3 x \( \ti{\ph} \pa_{\al} \ti{\ph}' - \ti{\ph}' \pa_{\al} \ti{\ph} \) \vep^{\al\be\ga} \frac{\pa \cH}{\pa E_{\fa}^{\be}} \frac{\pa \cH}{\pa B^{\fa\ga}} \propto \int \ed^3 x \( \ti{\ph} \pa_{\al} \ti{\ph}' - \ti{\ph}' \pa_{\al} \ti{\ph} \) q^{\al\be} \cD'_{\be} = 0 \, ,
\eeq
for some inverse 3-metric density $q^{\al\be}(E,B)$. Indeed, for the generic theory \eqref{eq:Scangen} all constraints correspond to gauge symmetries, so by diffeomorphism invariance we must have the closure $\{ \cH, \cH \} \sim \cD'$. By continuity in parameter space, this then also holds for the special theories considered here.

The seven equations \eqref{eq:dotC} and \eqref{eq:dotH} must therefore be satisfied for our constraints to be conserved. If $W^{\al\be,\ga\de} = 0$, then \eqref{eq:dotC} leads to six secondary constraints $\cC'^{\al\be} = 0$, since $\ti{\ph} \neq 0$ plays the role of the (densitized) lapse function. Along with the six primary constraints $\cC^{\al\be} = 0$, these reduce the canonical DoF count from 16 down to 4, i.e. the amount corresponding to a massless graviton. This is what happens in the case of GR and Self-Dual Gravity, because then $\cC^{\al\be}$ depends on only one of the two canonical fields and therefore commutes trivially with itself. In the case of the two new theories considered here, however, we have the same non-zero result
\beq \label{eq:Sidef}
W^{\al\be,\ga\de} = \al^2 \[ \vep^{\al\ga\ep} S_{\ep}^{\,\,\,\be\de} + \vep^{\al\de\ep} S_{\ep}^{\,\,\,\be\ga} + \vep^{\be\ga\ep} S_{\ep}^{\,\,\,\al\de} + \vep^{\be\de\ep} S_{\ep}^{\,\,\,\al\ga} \] \, ,
\eeq
where
\beq \label{eq:SOdef}
S_{\ga}^{\,\,\,\al\be} := B^{\fa(\al} \na_{\ga} E_{\fa}^{\be)} - E_{\fa}^{(\al} \na_{\ga} B^{\fa\be)} = - E_{\fa}^{\al} E_{\fb}^{\be} \na_{\ga} \psi^{\fa\fb}  \, .
\eeq
Note that this tensor density cannot be set to zero by lower-rank constraints, except for its trace
\beq
S_{\be}^{\,\,\,\al\be} = - E_{\fa}^{\al} E_{\fb}^{\be} \na_{\be} \psi^{\fa\fb} = - E_{\fa}^{\al} \na_{\be} \( \psi^{\fa\fb} E_{\fb}^{\be} \) = - E_{\fa}^{\al} \na_{\be} B^{\fa\be} \equiv 0 \, . 
\eeq
Because $W^{\al\be,\ga\de} \neq 0$, \eqref{eq:dotC} becomes a linear algebraic equation to solve for the 6-dimensional vector $\la_{\al\be}$. To analyse this equation we first express $W^{\al\be,\ga\de}$ as an operator by using the invertible metric density $E_{\al\be}$
\beq \label{eq:Wop}
\cW_{\al\be}^{\,\,\,\,\,\,\ga\de} := E_{\al\ep} E_{\be\ze} W^{\ep\ze,\ga\de} \, ,
\eeq
so that we can use the trace expression for $\cW$'s determinant. By explicit computation, we then note the useful property 
\beq \label{eq:TroddW}
\Tr \[ \cW^{2k+1 < 6} \] = 0 \, , 
\eeq
which implies in particular the following form for the set of eigenvalues
\beq \label{eq:specW}
{\rm spec}(\cW) = \{ \cW_1, \cW_2, \cW_3, - \cW_1, - \cW_2, -\cW_3 \} \, .
\eeq
With this the determinant reads
\beq
\det \cW = \frac{1}{8} \, \Tr \[ \cW^2 \] \Tr \[ \cW^4 \] - \frac{1}{48}\, \Tr \[ \cW^2 \]^3 - \frac{1}{6}\, \Tr \[ \cW^6 \] \, ,
\eeq
and the explicit result is a combination of all possible contractions of six $S_{\ga}^{\,\,\,\al\be}$ tensor densities, where the indices are displaced using $E_{\al\be}$ and its inverse. What matters here is that $\det \cW \neq 0$ for {\it generic} points of phase space. Thus, {\it generically} $\cW$ is an invertible matrix that determines completely $\la_{\al\be}$ in terms of the canonical fields
\beq \label{eq:laalbesol}
\la_{\al\be} = - \ti{\ph} Q_{\al\be,\ga\de} \cC'^{\ga\de} \, , \hspace{1cm} Q_{\al\be,\ep\ze} W^{\ep\ze,\ga\de} \equiv \de_{(\al}^{\ga} \de_{\be)}^{\de} \, . 
\eeq
Inserting this solution inside the action \eqref{eq:ScanC} then leads back to the generic form \eqref{eq:Scangen}, although with a modified Hamiltonian constraint  
\beq 
S = \int \ed^4 x \[ E_\fa^{\al} \dot{A}_{\al}^\fa - \ti{\ph} \cH' - N^{\al} \cD_{\al} - \te^\fa \cG_\fa \] \, , \hspace{1cm}  \cH' := \cH - Q_{\al\be,\ga\de} \, \cC^{\al\be} \cC'^{\ga\de} \, .
\eeq
Note that, as constraints, $\cH = 0$ and $\cH' = 0$ are equivalent, since $\cC^{\al\be} = 0$, but their variations are not and the modification $\cH \to \cH'$ is precisely such that $\cC^{\al\be} = 0$ is conserved \eqref{eq:dotC}. Also note that now $\cC^{\al\be} = 0$ has to be imposed by hand, since there is no longer an independent Lagrange multiplier to impose it through the variational principle, as usual for second-class constraints. 

Finally, we must also satisfy \eqref{eq:dotH}. Inserting \eqref{eq:laalbesol} we find
\beq \label{eq:QC2constr}
Q_{\al\be,\ga\de} \, \cC'^{\al\be} \cC'^{\ga\de} = 0 \, ,
\eeq
and note that, given $\cC^{\al\be} = 0$ and \eqref{eq:HHcom}, this is actually the self-commutator of the new Hamiltonian constraint
\beq
\{ H'[\ti{\ph}], H'[\ti{\ph}'] \} = 0 \, , \hspace{1cm} H'[\ti{\ph}] := \int \ed^3 x\, \ti{\ph} \cH' \, . 
\eeq
But this commutator has to be zero because of the local symmetries of the action, so we conclude that \eqref{eq:QC2constr} must hold automatically and therefore does not constitute a secondary constraint.\footnote{This is possible because the quadratic form $Q$ has the same signature as its inverse $W$ and thus $\cW$, whose signature is split \eqref{eq:specW}, so \eqref{eq:QC2constr} admits indeed non-trivial $\cC'^{\al\be} \neq 0$ solutions.} Having satisfied the conservation of all constraints, the Dirac algorithm terminates. The resulting canonical DoF freedom count is thus $16 - 6 = 10$, meaning that these theories {\it generically} have 5 DoF.

\subsection{Irregularity}

In the previous subsection we stressed that our conclusion holds for {\it generic} field values. In fact, it turns out that $W$ actually vanishes on the vacuum (A)dS solution \eqref{eq:MSsol}. To see this, first note that $W$ is entirely determined by the tensor density of rank three \eqref{eq:SOdef}. Given the spatial isometries of (A)dS, a rank-3 spatial tensor can only be essentially $\vep^{\al\be\ga}$, i.e. up to metric factors to displace indices and change the density weight. But \eqref{eq:SOdef} is symmetric in the two upper indices, so $\bar{W} = 0$. Next, we note that since (A)dS is a solution of all pure-connection theories, equation \eqref{eq:dotC} must trivially hold on that solution, which further implies $\bar{\cC}'^{\al\be} = 0$. The linear perturbation of \eqref{eq:dotC} around (A)dS is thus  
\beq
\bar{\ti{\ph}}\, \de \cC'^{\al\be} + \de W^{\al\be,\ga\de} \bar{\la}_{\ga\de} = 0 \, .
\eeq
This is independent of the Lagrange multiplier perturbations $\de \ti{\ph}$ and $\de \la_{\al\be}$ and therefore does not determine any of them to the order under consideration. Instead, it amounts to six secondary constraints for the linear fluctuations $\de A_{\al}^{\fa}$ and $\de E_{\fa}^{\al}$, thus leading to the degree of freedom count of GR, despite the fact that $W \neq 0$ when perturbations are included. In the case of the determinant theory, this is consistent with the fact we already mentioned that the linearisation of the action \eqref{eq:detS} around (A)dS is the same as the one of GR with zero Immirzi parameter.    

We have thus demonstrated the irregularity of our two candidate theories. On the one hand, we have seen that the linearised theories around (A)dS have less DoF than their fully non-linear counterparts, i.e. the phenomenon of field activation. On the other hand, the non-linear obstruction to secondary constraints comes from a field-dependent matrix $W$ that is generically non-degenerate, but is zero on (A)dS, meaning that we are vulnerable to constraint bifurcation. Note also that the field activation issue should also hold in the presence of matter, in which case the corresponding background solution would be the Friedmann-Lema\^itre-Robertson-Walker space-time, because all we needed to show $\bar{W} = 0$ where the spatial isometries of (A)dS.  

Finally, it is interesting to point out the special nature of the $\cC^{\al\be} = 0$ constraints for these peculiar theories. Usually one expects second-class primary constraints to generate secondary ones, so as to form a conjugate second-class pair (e.g. GR), since the reduced phase space dimension must be even. Instead, here we have that $\cC^{\al\be}$ becomes conjugate to itself, i.e. a conjugate pair of three second-class constraints. But this is possible only because $\cC^{\al\be}$ has an even number of independent components, which in turn is due to the number of spatial dimensions. In fact, even in phase space points where $W$ is degenerate $\det \cW = 0$, the resulting secondary constraints can only be of even number because of the spectrum form \eqref{eq:specW}, so the reduced dimension is again even. We therefore conclude that the present situation is specific to theories that can only be defined in certain dimensions. This is the case here, since $F \we F$ must be a form of maximal degree for the action \eqref{eq:Sf} to make sense.

\section{Conclusion} \label{sec:conclusion}

In this paper we have considered the broad class of diffeomorphism invariant gauge theories given by \eqref{eq:Sf}, where $X \sim F \we F$, and focused on the case where the gauge group is $G = {\rm SO}(1,3)$. GR is a member of that class, so this ``pure spin connection'' formulation offers an interesting alternative starting point for exploring modified theories of gravity. In particular, since these theories do not admit a perturbative expansion around Minkowski space-time, but only around (A)dS, they could in principle circumvent the standard uniqueness theorems of GR.

The generic theory of the considered class has eight DoF: a massless graviton, a massive graviton and a scalar, but one of the two gravitons is necessarily a ghost. Thus, only special theories with less DoF have a chance of being physically viable, as is the case of GR. The novel aspect of our work is the identification of a subset of theories, distinct from GR, which have less DoF than the generic member of the class, thanks to six extra primary constraints on phase space. In the case of GR, the conservation of these primary constraints leads to an equal number of secondary constraints, but this does not happen for the new theories reported here. As a result, these theories have five DoF, which lies exactly between the DoF count of GR and the generic theory. 

On the other hand, the linearisation of these theories around (A)dS has the same DoF count as GR, meaning that part of the DoF are activated only through interactions. This is a symptom of irregular dynamical systems and is due to the fact that the obstruction to secondary constraints depends on the phase space point. As argued in more detail in the Introduction, this situation is at best inconvenient, as it precludes the use of perturbative techniques, and at worst pathological, as it could still imply ghost-like instabilities or signal an ill-defined Cauchy problem.  It would be very interesting to understand the new theories discovered in this paper better. We leave this to future work.

\acknowledgments

The authors are grateful to Sergey Alexandrov for useful comments and suggestions. EM is supported by a Consolidator Grant of the European Research Council (ERC-2015-CoG grant 680886).

\appendix

\section{Determinant theory Hamiltonian constraint} \label{app:detpsi}

Here we compute the determinant of $\bm{\psi}$, given by \eqref{eq:psidettheory}, under the constraints \eqref{eq:EBnormal}
\bea
\det \bm{\psi} & \equiv & \frac{1}{6!}\, \vep_{\fa_1 \dots \fa_6} \vep_{\fb_1 \dots \fb_6} \, \psi^{\fa_1\fb_1} \dots \psi^{\fa_6 \fb_6} \nn \\
 & \equiv & \frac{1}{6!} \, \vep^{\fa_1 \dots \fa_6} \vep^{\fb_1 \dots \fb_6} E_{\al_1\be_1} \dots E_{\al_6 \be_6} \( E_{\fa_1}^{\al_1} B_{\fb_1}^{\be_1} + E_{\fb_1}^{\al_1} B_{\fa_1}^{\be_1} \) \dots \( E_{(\fa_6}^{\al_6} B_{\fb_6)}^{\be_6} + E_{\fb_6}^{\al_6} B_{\fa_6}^{\be_6} \)  \nn \\
 & \os{*}{\equiv} & \frac{1}{3!^2} \, \vep^{\fa_1 \dots \fa_6} \vep^{\fb_1 \dots \fb_6} E_{\al_1\be_1} \dots E_{\al_6 \be_6} \( E_{\fa_1}^{\al_1} B_{\fb_1}^{\be_1} \dots E_{\fa_3}^{\al_3} B_{\fb_3}^{\be_3} \) \( E_{\fb_4}^{\al_4} B_{\fa_4}^{\be_4} \dots E_{\fb_6}^{\al_6} B_{\fa_6}^{\be_6} \) \nn \\
 & \equiv & \frac{1}{3!^6} \, \vep^{\fa_1 \dots \fa_6} \vep^{\fb_1 \dots \fb_6} \vep^{\al_1\al_2\al_3} \vep^{\al_4\al_5\al_6} \vep^{\be_1\be_2\be_3} \vep^{\be_4\be_5\be_6} E_{\al_1\be_1} \dots E_{\al_6 \be_6}  \nn \\
 & & \times \( \vep_{\ga_1\ga_2\ga_3} E_{\fa_1}^{\ga_1} E_{\fa_2}^{\ga_2} E_{\fa_3}^{\ga_3} \) \( \vep_{\ga_4\ga_5\ga_6} B_{\fa_4}^{\ga_4} B_{\fa_5}^{\ga_5} B_{\fa_6}^{\ga_6} \) \( \vep_{\de_1\de_2\de_3} B_{\fb_1}^{\de_1} B_{\fb_2}^{\de_2} B_{\fb_3}^{\de_3} \) \( \vep_{\de_4\de_5\de_6} E_{\fb_4}^{\de_4} E_{\fb_5}^{\de_5} E_{\fb_6}^{\de_6} \) \nn \\
 & \equiv & \frac{1}{3!^4} \, \vep^{\fa_1 \dots \fa_6} \vep^{\fb_1 \dots \fb_6} E^{-2}  \nn \\
 & & \times \( \vep_{\ga_1\ga_2\ga_3} E_{\fa_1}^{\ga_1} E_{\fa_2}^{\ga_2} E_{\fa_3}^{\ga_3} \) \( \vep_{\ga_4\ga_5\ga_6} B_{\fa_4}^{\ga_4} B_{\fa_5}^{\ga_5} B_{\fa_6}^{\ga_6} \) \( \vep_{\de_1\de_2\de_3} B_{\fb_1}^{\de_1} B_{\fb_2}^{\de_2} B_{\fb_3}^{\de_3} \) \( \vep_{\de_4\de_5\de_6} E_{\fb_4}^{\de_4} E_{\fb_5}^{\de_5} E_{\fb_6}^{\de_6} \) \nn \\
 & \equiv & \frac{1}{3!^4} \, \vep^{\fa_1 \dots \fa_6} \vep^{\fb_1 \dots \fb_6} E^{-2} \( \vep_{\ga_1\ga_2\ga_3} E_{\fa_1}^{\ga_1} E_{\fa_2}^{\ga_2} E_{\fa_3}^{\ga_3} \) \( \vep_{\de_1\de_2\de_3} E_{\fb_1}^{\de_1} E_{\fb_2}^{\de_2} E_{\fb_3}^{\de_3} \)  \nn \\
 & & \times \( \vep_{\ga_4\ga_5\ga_6} B_{\fa_4}^{\ga_4} B_{\fa_5}^{\ga_5} B_{\fa_6}^{\ga_6} \)  \( \vep_{\de_4\de_5\de_6} B_{\fb_4}^{\de_4} B_{\fb_5}^{\de_5} B_{\fb_6}^{\de_6} \) \nn \\
 & \os{**}{=} & -\frac{1}{3!^4} \, E^{-2} \( \vep_{\ga_1\ga_2\ga_3} E_{\fa_1}^{\ga_1} E_{\fa_2}^{\ga_2} E_{\fa_3}^{\ga_3} \) \( \vep_{\de_1\de_2\de_3} E^{\fa_1\de_1} E^{\fa_2\de_2} E^{\fa_3\de_3} \)  \nn \\
 & & \times \( \vep_{\ga_4\ga_5\ga_6} B_{\fa_4}^{\ga_4} B_{\fa_5}^{\ga_5} B_{\fa_6}^{\ga_6} \)  \( \vep_{\de_4\de_5\de_6} B^{\fa_4\de_4} B^{\fa_5\de_5} B^{\fa_6\de_6} \)  \nn \\
 & \equiv & -\frac{1}{3!^4}\, E^{-2} \( \vep_{\ga_1\ga_2\ga_3} \vep_{\de_1\de_2\de_3} E^{\ga_1\de_1} E^{\ga_2\de_2} E^{\ga_3\de_3} \)  \( \vep_{\ga_4\ga_5\ga_6} \vep_{\de_4\de_5\de_6} B^{\ga_4\de_4} B^{\ga_5\de_5} B^{\ga_6\de_6} \)  \nn \\
 & \equiv & - E^{-1} B   \, .  
\eea
In the $*$ step we have used the fact that only a maximum of three $E_{\fa}^{\al}$ or $B_{\fa}^{\al}$ can be antisymmetrized, which thus leaves ${6 \choose 3}$ repetitions of the only surviving combination. In the $**$ step we have expressed the product of Levi-Civita tensors in terms of the antisymmetrized $\ka^{\fa\fb}$ product, taking into account that $\ka_{\fa\fb}$ has determinant $-1$ and the orthogonality constraint \eqref{eq:EBnormal}, so that there is only one non-trivial term in the sum.

\bibliographystyle{JHEP}
\bibliography{mybib}

\end{document}

%% file: mydefs.tex
\newcommand{\beq}{\begin{equation}}
\newcommand{\eeq}{\end{equation}}
\newcommand{\bea}{\begin{eqnarray}}
\newcommand{\eea}{\end{eqnarray}}
\newcommand{\ben}{\begin{enumerate}}
\newcommand{\een}{\end{enumerate}}

\newcommand{\pa}{\partial}

\newcommand{\na}{\nabla}

\newcommand{\ed}{{\rm d}}
\newcommand{\ced}{{\rm D}}
\newcommand{\we}{\wedge}

\newcommand{\Ord}{{\cal O}}

\newcommand{\Tr}{{\rm Tr}}
\newcommand{\ti}{\tilde}

\newcommand{\Rs}{\mathbb{R}}
\newcommand{\Cs}{\mathbb{C}}

\renewcommand\({\left(}
\renewcommand\){\right)}
\renewcommand\[{\left[}
\renewcommand\]{\right]}

\newcommand{\os}{\overset}

\newcommand{\nn}{\nonumber}

\newcommand{\al}{\alpha}
\newcommand{\be}{\beta}
\newcommand{\ga}{\gamma}

\newcommand{\de}{\delta}

\newcommand{\ep}{\epsilon}
\newcommand{\vep}{\varepsilon}
\newcommand{\ze}{\zeta}
\newcommand{\et}{\eta}
\newcommand{\te}{\theta}

\newcommand{\ka}{\kappa}
\newcommand{\la}{\lambda}
\newcommand{\La}{\Lambda}
\newcommand{\ro}{\rho}
\newcommand{\si}{\sigma}
\newcommand{\Si}{\Sigma}

\newcommand{\ph}{\phi}

\newcommand{\cH}{{\cal H}}

\newcommand{\cA}{{\cal A}}

\newcommand{\cD}{{\cal D}}

\newcommand{\cC}{{\cal C}}
\newcommand{\cG}{{\cal G}}

\newcommand{\cF}{{\cal F}}
\newcommand{\cW}{{\cal W}}

\newcommand{\fa}{{\frak{a}}}
\newcommand{\fb}{{\frak{b}}}
\newcommand{\fc}{{\frak{c}}}
\newcommand{\fd}{{\frak{d}}}